\documentclass[prd,aps,twocolumn,floats,nofootinbib, showpacs,floatfix,superscriptaddress]{revtex4-1}

\usepackage{graphicx}
\usepackage{epstopdf}
\usepackage{url}
\usepackage[colorlinks=true, citecolor=blue, urlcolor = blue, linkcolor= blue, bookmarks=true]{hyperref}
\usepackage{color}
\usepackage{amsfonts}

\begin{document}

\title{\protect Computational Orbital Mechanics of Marble Motion on a 3D Printed Surface 
--  \\1. Formal Basis}

\author{Pooja Bhambhu
}
\affiliation{D. S. Kothari Center for Research and Innovation in Science Education, Miranda House, University of Delhi, Delhi $-$ 110 007, India.}
\affiliation{Department of Physics, Miranda House, University of Delhi, Delhi $-$ 110 007, India.}

\author{Preety 
}
\affiliation{D. S. Kothari Center for Research and Innovation in Science Education, Miranda House, University of Delhi, Delhi $-$ 110 007, India.}
\affiliation{Department of Physics, Miranda House, University of Delhi, Delhi $-$ 110 007, India.}
\affiliation{Department of Physics, Indian Institute of Technology Bombay, Powai, Mumbai $-$ 400076, India.}

\author{Paridhi Goel}
\affiliation{D. S. Kothari Center for Research and Innovation in Science Education, Miranda House, University of Delhi, Delhi $-$ 110 007, India.}
\affiliation{Department of Physics, Miranda House, University of Delhi, Delhi $-$ 110 007, India.}
\affiliation{Department of Physics, Indian Institute of Technology Delhi, Hauz Khas, Delhi $-$ 110016, India.}

\author{Chinkey 
}
\affiliation{D. S. Kothari Center for Research and Innovation in Science Education, Miranda House, University of Delhi, Delhi $-$ 110 007, India.}
\affiliation{Department of Physics, Miranda House, University of Delhi, Delhi $-$ 110 007, India.}

\author{Manisha Siwach}
\affiliation{D. S. Kothari Center for Research and Innovation in Science Education, Miranda House, University of Delhi, Delhi $-$ 110 007, India.}
\affiliation{Department of Physics, Miranda House, University of Delhi, Delhi $-$ 110 007, India.}

\author{Ananya Kumari}
\affiliation{D. S. Kothari Center for Research and Innovation in Science Education, Miranda House, University of Delhi, Delhi $-$ 110 007, India.}
\affiliation{Department of Physics, Miranda House, University of Delhi, Delhi $-$ 110 007, India.}

\author{Sudarshana 
}
\affiliation{D. S. Kothari Center for Research and Innovation in Science Education, Miranda House, University of Delhi, Delhi $-$ 110 007, India.}
\affiliation{Department of Physics, Miranda House, University of Delhi, Delhi $-$ 110 007, India.}

\author{Sanjana Yadav}
\affiliation{D. S. Kothari Center for Research and Innovation in Science Education, Miranda House, University of Delhi, Delhi $-$ 110 007, India.}
\affiliation{Department of Physics, Miranda House, University of Delhi, Delhi $-$ 110 007, India.}

\author{Shikha Yadav}
\affiliation{D. S. Kothari Center for Research and Innovation in Science Education, Miranda House, University of Delhi, Delhi $-$ 110 007, India.}
\affiliation{Department of Physics, Miranda House, University of Delhi, Delhi $-$ 110 007, India.}

\author{Bharti 
}
\affiliation{D. S. Kothari Center for Research and Innovation in Science Education, Miranda House, University of Delhi, Delhi $-$ 110 007, India.}
\affiliation{Department of Physics, Govt. P. G. College, Hisar $-$ 125001, Haryana, India.}

\author{Poonam 
}
\affiliation{D. S. Kothari Center for Research and Innovation in Science Education, Miranda House, University of Delhi, Delhi $-$ 110 007, India.}
\affiliation{Department of Physics, Govt. College for Girls, Hisar $-$ 125001, Haryana, India.}

\author{Anshumali}
\affiliation{D. S. Kothari Center for Research and Innovation in Science Education, Miranda House, University of Delhi, Delhi $-$ 110 007, India.}
\affiliation{Department of Physics, Miranda House, University of Delhi, Delhi $-$ 110 007, India.}

\author{Athira Vijayan}
\affiliation{D. S. Kothari Center for Research and Innovation in Science Education, Miranda House, University of Delhi, Delhi $-$ 110 007, India.}
\affiliation{Department of Physics, Miranda House, University of Delhi, Delhi $-$ 110 007, India.}

\author{Divakar Pathak}
\email{divakar.pathak@mirandahouse.ac.in, dpdlin@gmail.com}
\affiliation{D. S. Kothari Center for Research and Innovation in Science Education, Miranda House, University of Delhi, Delhi $-$ 110 007, India.}
\affiliation{Department of Physics, Miranda House, University of Delhi, Delhi $-$ 110 007, India.}

\begin{abstract}
Simulating curvature due to gravity through warped surfaces is a common visualization aid in Physics education. We reprise a recent experiment exploring orbital trajectories on a precise 3D-printed surface to mimic Newtonian gravity, and elevate this analogy past the status of a mere visualization tool. We present a general analysis approach through which this straightforward experiment can be used to create a 
reasonably 
advanced computational orbital mechanics lab at the undergraduate level, creating a convenient hands-on, computational pathway to various non-trivial nuances in this discipline, such as the mean, eccentric, and true anomalies and their computation, Laplace-Runge-Lenz vector conservation, characterization of general orbits, and the extraction of orbital parameters. We show that while the motion of a marble on such a surface does not truly represent a orbital trajectory under Newtonian gravity in a strict theoretical sense, but through a proposed projection procedure, the experimentally recorded trajectories 
closely resemble the Kepler orbits and 
approximately respect the known conservation laws for orbital motion. 
The latter fact is demonstrated through multiple experimentally-recorded elliptical trajectories with wide-ranging eccentricities and semi-major axes.  

In this first part of this two-part sequence, we lay down the formal basis of this exposition, describing the experiment, its calibration, critical assessment of the results, and the computational procedures for the transformation of raw experimental data into a form useful for orbital analysis.

\end{abstract}

\maketitle

\begin{section}{Introduction}\label{introduction}

The advent of additive manufacturing technology, more commonly known as the 3D printing technology, has had an immensely transformational value in diverse fields of science and technology \cite{3DPrint_PhysToday}. Its ubiquity has progressively grown in recent years, with multiple disciplines involving manufacturing and design reaping the benefits of the control and maneuverability this technology allows \cite{ICTP_Lowcost3Dprint}. In particular, the introduction of this technology into modern pedagogy has a potential to revolutionize science education \cite{Reiss, Segerman_maths_3dprinting, Knill_arXiv, Abu}, and to the extent it has been introduced thus far, it has already begun to have a very constructive impact \cite{ICTP_Lowcost3Dprint, Reiss, 3DPrint_MNRAS}. 
Among the myriad ways in which science education can benefit from this technology, one fairly recent example stands out in both depth and scope. In mathematically-extensive branches of Physics such as the theory of gravitation/general relativity, and orbital mechanics, a common visualization aid traditionally adopted in classrooms is to use the motion of ball on a warped spandex fiber to mimic the motion of a massive body under the gravitational influence of another. It is convenient to draw this analogy as a pedagogic tool to help students visualize how a massive body can warp space, and physically motivate the idea of curvature used often abstractly in a course on general relativity. This analogy, and its impreciseness has been the subject of many theoretical and experimental investigations over the years, such as Refs. \cite{Middleton, Middleton2, Whitewalker_Spandex_shape_Orbits, Nauenberg} (and the references therein, these are merely representative examples).  In this regard, the use of a 3D printer to print an exact replica of an appropriately designed mathematical surface is a complete game-changer. It renders far larger amount of control to the experimenter than was ever traditionally available in spandex-based works preceding this 3D-printing era. As we witness in the coming sections, with the aid of 3D printing technology, one is liberated from the restriction of having to perform selective examination/analysis of low-eccentricity, near-circular orbits, or a selective analysis of specific regions of the warped-surface having high or low curvature. The technology makes it possible to easily observe and analyze truly 3D motion, in an elliptic arc/orbit of any eccentricity whatsoever. 

The key idea underlying such pedagogic improvement is fairly straightforward. Since the gravitational potential energy of any point mass $m$ located a certain $z-$ distance above a reference mark is $U = m g z$, and force is related to this $U$ as its negative gradient \cite{Goldstein}, if the point mass in question is undergoing motion on a surface where the $z-$ coordinate scales with the radial distance from the center, $r$, as $z = 1/r$, the negative radial gradient of $U$ can generate the familiar Newtonian gravitational force $F \propto -1/r^2$. Through their pioneering work, Lu et al. \cite{PhysEd3DPrint, AmJPhys3DPrint} demonstrated that by launching marbles on a surface having this precise mathematical form, it is possible to visualize Kepler's laws of planetary motion for low-eccentricity, near-circular orbits. But strictly speaking, one can easily infer that this analogy is mathematically imprecise since such geometries/potentials only permit cylindrical symmetry about a axis through the center, rather than the complete spherical symmetry of the Newtonian gravitational potential.
An elaborate mathematical analysis \cite{Middleton, Middleton2} of these two distinct physical problems brings to the fore some key points of difference. The Lagrangians, as well as the ensuing equations of motion in the two cases are distinct, and in general, there exists no spherically symmetric potential which generates the precise radial derivative needed for making these problems equivalent. Thus, in general, no cylindrically symmetric surface in uniform gravity can reproduce the exact motion of a body in any spherically symmetric potential, except in the special case of perfectly circular orbits. Also, on cylindrically symmetric surfaces, the orbital trajectories feature precessional motion, due to which their mathematical form may be approximated  \cite{Middleton} as
\begin{equation}
r(\phi) = r_0 \left( 1 - \epsilon \cos (\nu \phi) \right),
\end{equation}
where $\nu$ refers to the precession frequency. These may be contrasted against the usual Kepler orbits \cite{Goldstein}, which are non-precessing (stationary) and correspond to the special case of $\nu = 1$. This subset of solutions is also permitted in cylindrical geometries, but the possibility is a little restricted -- only for specific curvature of surface. A natural consequence of precession of apocenters also is that the such orbital trajectories will not appear closed under one orbital motion cycle, which entails the violation of Kepler's first law in this context, as well as a non-conservation of angular momentum in very strict terms. The extent of the violation/non-conservation depends on the extent of deviation from perfect stationary orbits, or the extent to which $\nu$ differs from unity. 

In this context, it is only natural to wonder, what exactly is to be defined as the goal of this experiment. One plausible direction of investigation \cite{AmJPhys3DPrint} can be to systematically probe these analytic theoretical predictions \cite{Middleton} by printing surfaces of different curvatures. But pedagogically, it is far more appealing that in spite of the fundamental difference between the two geometries, Lu et al. were able to demonstrate \cite{PhysEd3DPrint} the approximate validity of the Keplerian analogy by studying circular/low-eccentricity orbits, which exhibits both the utility of this simple experiment as a significant visualization aid, and also the power of 3D-printing technology in this context. 

Both these perspectives are plausible, but as neutral third-party readers of the original articles, our beliefs align with the latter perspective. 
With all this emphasis on the exactness of the solutions, precession, and the comparison of the actual observed marble motion against the predictions of the theory, it is easy to lose track that our primary starting goal in this experiment was not to provide a testing ground to gauge how accurate are the predictions of the theory vis-a-vis the actual motion of the marble, but rather, to provide a pedagogic visualization aid, to assist the exposition of Kepler's laws to students by giving it an experimental basis. While this analogy may not be perfect, this is still a well crafted visualization aid, and an immense pedagogic simplification. If we weigh the pros and cons, the immense pedagogic benefits outweigh the inexactness/inaccuracy of the solutions, as is indeed the case in many physical examples. In Physics pedagogy, such compromises are not unheard of: we don't consider the light emanating from optical sources as the mathematically exact wave packets, but rather, as the inexact but pedagogically-more-useful infinite sinusoidal wave trains. Likewise, clouds may not be spheres and mountains may not be cones, but the actual pedagogic cost of treating a mountain as a recursive fractal improvement over a cone/triangle, is some knowledge of geometry inaccessible to a middle school student. Often times, a mathematically inexact/oversimplified model, is pedagogically more beneficial than a complicated, exact, mathematical solution. 

We reason that precession is only relevant when multiple orbital cycles are completed, which is hardly plausible in the presence of friction. For the ellipse-like orbital trajectories one is likely to observe in such experiments, anything in excess of one orbital cycle is atypical. We thus propose to neglect the precession of orbits, assuming $\nu = 1$, which is not far removed from the experimental observations of Lu et al. \cite{AmJPhys3DPrint}, who also found $\nu$ to be only marginally larger. 
Neglecting the precession of apocenters reduces these general trajectories to the smaller subset of Keplerian, stationary orbits analyzed in \cite{Middleton}, which is not only an immense simplification, but also precisely what we are aiming to simulate through this experiment. 

With the spirit of tapping into the complete potential of this visualization aid in the larger context, in the present work, we extend the scope of the above experiment, well past the visualization-aid and theory-validation perspectives hitherto considered. We illustrate how the same experiment presents a wonderful opportunity to develop a full-blown  computational orbital mechanics lab, which differs from the traditional approach of numerical lab exposure in orbital mechanics \cite{Curtis}, in that, instead of sourcing the data from some third-party observations, the source of the data here is an experiment performed by the students themselves, hence making it a sufficiently `hands-on' lab catering to both computation and experiment. Reporting on the exercises performed in an undergraduate lab setting, we additionally demonstrate that the fact that in a strict mathematical sense, the marble motion on a 3D printed surface is not exactly equivalent to motion of planetary bodies in Keplerian orbits, is a great blessing-in-disguise, as it can be turned into an opportunity to introduce some numerical techniques in significantly more applied context, hence enriching student understanding of numerous facets of linear algebra, conic sections, and curve-fitting, in addition to the main theme of the experiment -- orbital mechanics.  

\end{section}

\begin{section}{The Experiment}

We reprise the original experiment of Lu et al. \cite{PhysEd3DPrint, AmJPhys3DPrint}, with some slight modifications. Fig. \ref{3Dprintevolimage} illustrates various stages of surface design and print. The desired $z = 1/r$ surface can mathematically plotted through a simple code \cite{Github_OrbM} and subsequently exported in the STL format which can be read and recognized by a 3D printer. Since this experiment 
requires the motion of the marble to occur on the said surface, instead of preparing a laminar surface like Ref. \cite{AmJPhys3DPrint}, we thicken the region underlying the top surface with an additional surface fill to better support the weight of the marble. This does not affect the mathematical premise of the problem, but allows for greater flexibility in marble choice, since a laminar surface can, in principle, topple/shake 
with a very heavy marble, thereby affecting the observed orbital trajectories. Also, rather than print $z = -1/r$ surface \cite{PhysEd3DPrint, AmJPhys3DPrint}, we work 
with a $z = 1/r$ surface with underlying surface fill, inverted manually, which is mathematically equivalent, and is shown in Fig. \ref{3Dprintevolimage}(a). The exported STL file, shown in Fig. \ref{3Dprintevolimage}(b), is then scaled to larger dimensions while preserving the aspect ratio to preserve the surface shape, and 
is sliced to generate a Gcode file which can be used by a Colido $2.0 +$ printer for printing. This surface is printed with the filament material made of the thermoplastic 
Acrylonitrile Butadiene Styrene (ABS), and has physical dimensions of 
15.5 cm $\times$ 15.5 cm laterally, 
with a depth of $4.3$ cm from the center of top-surface, and $4.756$ cm from the topmost (diagonal) edge, as shown in Fig. \ref{3Dprintevolimage}(c). From the equation, $r \rightarrow 0$ only when the depth is infinite, but
since we are printing only a finite region of the whole inverse-$r$ surface, a finite depth value ensures that we don't actually reach $r \rightarrow 0$ in the lower surface. Rather, the surface carves out a circle of radius $R_{\rm in} = 2$ cm at the bottom surface of the sample, as can be physically measured.
This printed surface is reasonably fine in terms of finish and texture, and requires no external polishing (unlike Ref. \cite{AmJPhys3DPrint}), but since the very process of 3D printing entails layer-by-layer printing with a finite tip size, howsoever small the tip may be, the final structure is bound to have a staircase texture on close examination. This is inevitable but presents no major experimental difficulty. Since any external smoothing/scrubbing can potentially dent the surface, causing a deviation from the precise mathematical form, we prefer to circumvent this problem by purposefully choosing large, relatively heavier marbles (with a typical radius {\color{black} $\sim 0.5$ cm)}, whose motion on the surface is observed to be smooth even within the limits of computational accuracy of this experiment. We find this choice and particularly, the finite size, to not at all be restrictive with regard to the ensuing analysis, since any spherical marble makes contact with a rigid surface at a single point only, and so, while tracking this marble motion, we regard just a single point at the center of the marble moving smoothly on the surface. 

\begin{figure}
\begin{center}
\scalebox{0.6}{\includegraphics{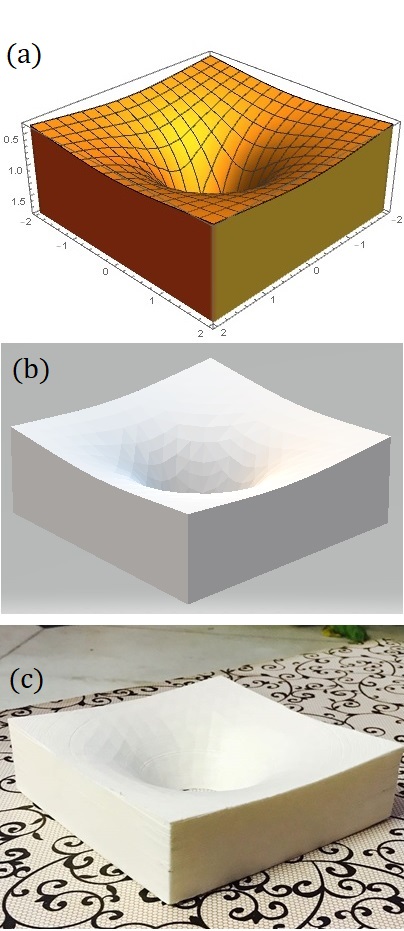}}\caption{ \label{3Dprintevolimage} (Color Online) Generation of the 3D print of the $z = 1/r$ surface used in this work. The figure depicts the three stages: (a) The Mathematica surface plot of the said surface, with additional surface fill beneath the desired surface for support. (b) The corresponding STL file, which gets fed into the 3D printer. (c) The actual surface output obtained from the 3D printer, with lateral dimensions 15.5 cm squared.}
\end{center}
\end{figure}

\begin{figure}
\begin{center}
\scalebox{0.7}{\includegraphics{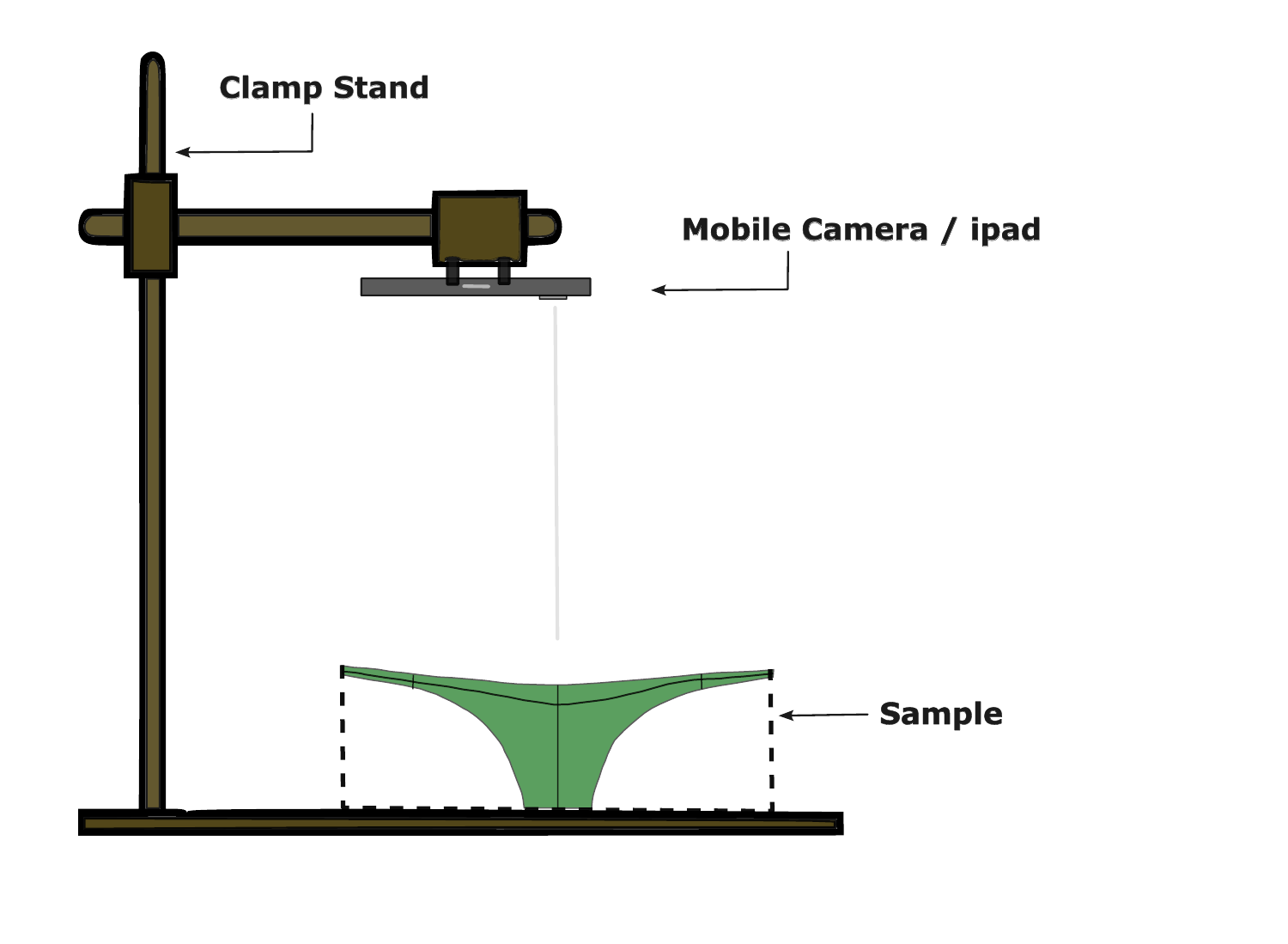}}\caption{ \label{ExptSetup} (Color Online) A schematic diagram of the experimental setup used. For the clarity of premise, only the inner surface is annotated in this schematic; the complete sample region with the underlying fill is represented through dotted lines. The mobile/iPad camera is clamped in position some vertical distance above the 3D printed surface, and records the trajectories followed by marbles launched on the surface.}
\end{center}
\end{figure}

A schematic diagram of the actual experimental set up is shown in Fig. \ref{ExptSetup}. The recording camera is clamped in position some vertical distance above the 3D printed surface, and records the trajectories followed by marbles launched on the surface. Many aspects of this experiment are purposefully kept simple to ensure that the experiment is more accessible to undergraduate students. Marbles were manually launched approximately tangentially on the surface. Only a subset of the attempts resulted in Kepler-esque orbital trajectories, as sometimes, due to manual variability in the launch velocity, the motion of marble was observed to be radially spiral (velocity not entirely tangential), or direct radial descent (insufficient velocity). Nevertheless, with a large number of trials, sufficiently large number of viable data sets could be recorded, some of which, along with the ensuing computational analysis, are collected in a companion website \cite{OrbM_website} to this article, which we maintain. For recording the orbital trajectories, we used simple mobile phone cameras, as well as an iPad, which recorded videos of 30 and 120 frames per second (fps). Naturally, the time-resolution (time between frames) of the data sampling is the inverse of the fps value of the recording device (e.g. $\Delta t = 1/30$ s in the former case), and is an important parameter affecting the computational accuracy, since e.g. velocity may be approximated as a finite difference of the instantaneous coordinates over this very time-resolution. While ideally, a higher fps value is desirable for better accuracy, we find even $30$ fps to be sufficient for our current purposes. These recorded videos were analyzed by importing them into the popular video-analysis free-ware Tracker \cite{Tracker}. A powerful feature of Tracker, called the calibration stick, allows one to map coordinate axes on the recorded video and scale these to the known physical distances, so how much exactly is the actual distance between the camera and the sample is irrelevant. From the top-normal-view, the camera can only record the lateral position at periodic intervals of time, and hence this Tracker-analysis provides just the $x$ and $y$ coordinates for orbital motion at different times. The depth information ($z$ coordinate) is not directly recorded in the video, but can be calculated from the $x$ and $y$ coordinates recorded by invoking the equation of the surface after calibration. All subsequent computational analysis on this coordinate-time data extracted from Tracker, can be performed by importing these into the open-source software for numerical computation, Scilab \cite{Scilab_site}.

\end{section}

\begin{section}{Calibration}

\subsection{Mathematical Model}

Since the depth information ($z$-coordinate) in this experiment is to be obtained from the equation of the surface using the $x$ and $y$ coordinates extracted from Tracker, the knowledge of the exact mathematical form of the printed surface is imperative.
However, determining the same is a slightly non-trivial issue, as one may reason from simple dimensional considerations. This $z = 1/r$ surface designed as above, must actually be of the form $z = \kappa/r$, where $\kappa$ has dimensions of area, and is just numerically equal to 1 in our surface specification at the beginning. In order to avoid distorting the shape of the sample while scaling the prototype of the 3D-print to larger dimensions, an important precaution we observed was to lock the aspect ratio so that all three coordinates scale equally through some $L_{\rm scale}$ factor. But even when lengths scale equally, area would scale as $L_{\rm scale}^2$, so that $z = \kappa/r$ correctly scales by $L_{\rm scale}$ itself, hence preserving the aspect ratio as it should. However, on careful reflection, this also entails that on scaling the prototype, while the surface still remains a $(1/r)$ surface, the mathematical form of the surface does not remain the same $z = \kappa/r$ (with the original value of $\kappa = 1$) due to unequal scaling of $\kappa$ and $r$ in this expression. 
Therefore, 
we model this printed surface to be 
mathematically represented 
as 
$z = C_1 +  
(C_2/r) $, 
where $C_2 = - \vert C_2 \vert$ is a different constant than the original $\kappa$, and $C_1$ is another constant introduced owing to spatial translation due to the adoption of any particular choice of origin, or coordinate frame. These constants can be evaluated in two ways: mathematically, using sample geometry as an input, and experimentally, through a sampling of the surface heights against a calibrated horizontal axis. We describe the latter (experimental) approach in the next section. As for the former approach, we adopt a coordinate system shown in Fig. \ref{3dprinttheoryannotations} and choose $z = 0$ at the highest elevated points of the surface, which are the four diagonal vertices of the sample. With reference to this choice of coordinates, the constants can be evaluated to be:
\begin{equation}
C_2 = - \vert z_{\rm depth}\vert \frac{R_{\rm out} \ R_{\rm in}}{\left( R_{\rm out} -R_{\rm in} \right)}
\end{equation}
\begin{equation}
C_1 = - \frac{C_2}{R_{\rm out}} = \vert z_{\rm depth}\vert \frac{R_{\rm in}}{\left( R_{\rm out} -R_{\rm in} \right)}
\end{equation}
Substituting the values of $R_{\rm in} = 2$ cm, and $R_{\rm out} = 15.5/\sqrt{2} = 10.96$ cm, and $z_{\rm depth} = 4.756$ cm, we obtain the following theoretical estimates for the surface constants.
\begin{equation}
C_1 = 1.0616 \ {\rm cm}, \ {\rm and} \ C_2 = -11.6357 \ {\rm cm}^2
\end{equation}
Complete details of the calculation can be found in the Supplementary Information \cite{Supplement} to this article. 


\begin{figure}
\begin{center}
\scalebox{0.57}{\includegraphics{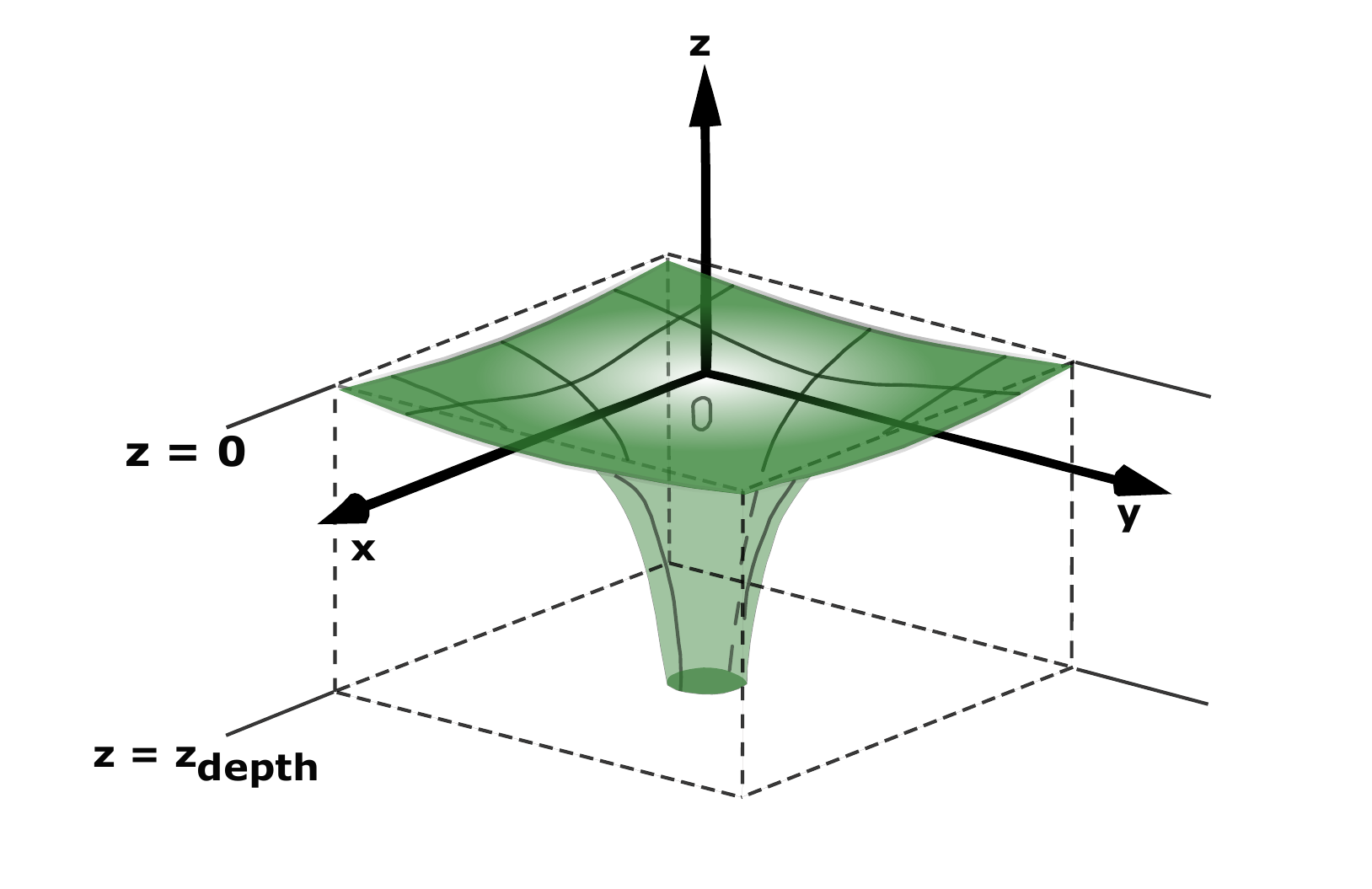}}\caption{ \label{3dprinttheoryannotations} (Color Online) The geometry and coordinate choice for surface calibration. As before, for clarity, only the inner surface is shown; the complete sample region with the underlying fill is represented as dotted lines. $z=0$ is chosen at the highest elevated point of the sample, corresponding to the diagonal edges. At the bottom surface $(z = z_{\rm depth} = - \vert z_{\rm depth} \vert)$, the surface carves out a circle of radius $R_{\rm in}$.}
\end{center}
\end{figure}

\subsection{Experimental Calibration}

The experimental surface-calibration process requires us to sample the surface at different points in the $x-z$ plane, and to fit a curve through these points. However, this requires a cross sectional view of the inner surface of the sample, which is not accessible for direct experimentation. We adapt the experimental arrangement to provide for the same, by adding an additional reference mark provided by a pin-thermocol arrangement placed underneath the sample, which allows us to circumvent these difficulties and perform the calibration experimentally. We refer the interested reader to the Supplementary Information \cite{Supplement} to this article, where complete details about the experimental assembly have been collected. After collecting various data points/sets as desired, we best fit the data to the expected curve to determine the values of the constants. This can be done through multiple procedures, all of them mutually equivalent:
\begin{enumerate}
\item Writing a simple routine for least squares fitting in any programming language/software package, upon importing this experimental $(x_i, z_i')$ data. 
\item Importing this data in Scilab \cite{Scilab_site}, and using the built-in \emph{reglin} subroutine for linear regression, directly yielding the surface constants. 
\item Simply entering this experimental $(x_i, z_i')$ data in Microsoft Excel, and using Excel Solver to find an optimal fit. This can be easily done for both $z'$-versus-$(1/x)$, as well as for $z'$-versus-$x$ variation. The procedure requires some initial guesses for $C_2$ and $C_3$ to be provided, for which the theoretical estimates obtained earlier  prove very useful. The Solver optimizer searches for optimal regression solutions in the vicinity of the guesses we provide, and typically returns surface constants' values only slightly different from the theoretical values obtained earlier. 
\end{enumerate}

We find the experimental values of the surface constants to be as follows: 
\begin{equation}
C_2 = -12.0717 \ {\rm cm}^2, \ \ \ \ C_1 = \vert = 1.0646 \ {\rm cm}
\end{equation}
Thus, we find that the theoretical and experimental values of these surface constants are in reasonably good agreement. The experimental value of $C_3$ in particular is in excellent agreement with the theoretical estimate, while the experimental value of $C_2$ is also found to be within $4 \%$ of the theoretical estimate. This agreement lends a lot of credence to this model formulated. 


\end{section}

\begin{section}{Critical Assessment of Kepler's First Law for Marble Motion}
\label{Critical_Assessment_KeplerFirstLaw_section}

As discussed in Section 1, whether or not the marble motion preserves Kepler's first law is a key question, which can now be simply answered on the basis of the data recorded from this experiment. For this, one can use a simple trick from vector analysis. One can use just three consecutive data points from the experimentally recorded datasets, two compute two consecutive distance vectors whose cross product can be used to identify the normal vector to the plane constituted by them.  If the points are distributed in a plane, the normal vectors so identified will be parallel to one another, and vice-versa. Moreover, from the information available, one can also compute the angles between these normal vectors to gauge the extent of coplanarity of these points. 

The typical results are shown in Figs. 4 -- 6, showing respectively the typical orbital trajectories (for an elliptical-like, closed) orbit, the distribution of the normal vectors to the triplets of consecutive data points, and the distribution of the angles between consecutive normal vectors. Fig. 4 seems to suggest that the motion of points is approximately in a plane to a very good approximation, but Figs. 5 and 6 suggest that there is indeed a small variation from complete planarity. The largest deviation angle is $\sim 30$ for a single data point, but the typical deviation angles range around $< 10^{\circ}$, which is significantly small. But howsoever small this angle of deviation may be, in a strict mathematical sense, this is still an explicit experimental verification of the theoretical argument \cite{Middleton} for this motion of the ball to be distinct from Keplerian orbits.

However, the results of Fig. 4 are still reassuring in that one can see an approximate restoration of the initial coordinates at the end of the orbit. Considering that we are using a finite sized marble, we can envisage in general two kinds of orbit closures - a strong form and a weak form. A strict ``fully-closed'' orbit would mean the complete restoration of the initial coordinates $(x, y, z)$ at the end of the orbit, a very strict condition which upon a large number of trials, we do occasionally observe to be satisfied within a single decimal place or two. A majority of the trials result in elliptical arcs, rather than these fully closed orbits, which is understandable since the dissipative effects of friction can not be completely eliminated in this experiment. However, we can also have an intermediate subset of elliptical trajectories, characterized henceforth as ``near-closed'' orbits, which we define as pertaining to the condition $\left( {\vec r_f} - {\vec r_i} \right) < R_{\rm ball}$, or even some percentage of $R_{\rm ball}$. This generous definition accommodates a much bigger subset of the observed elliptical trajectories, in the category of closed orbits. While not very rigorous, this still serves our purpose very well in the present work, since ultimately, one requires merely five points on an ellipse to uniquely determine it \cite{Anton, MSE_5pointsdetermineellipse}, so in principle, each of these elliptical trajectories is actually an overdetermined system.

\begin{figure}
\begin{center}
\scalebox{0.4}{\includegraphics{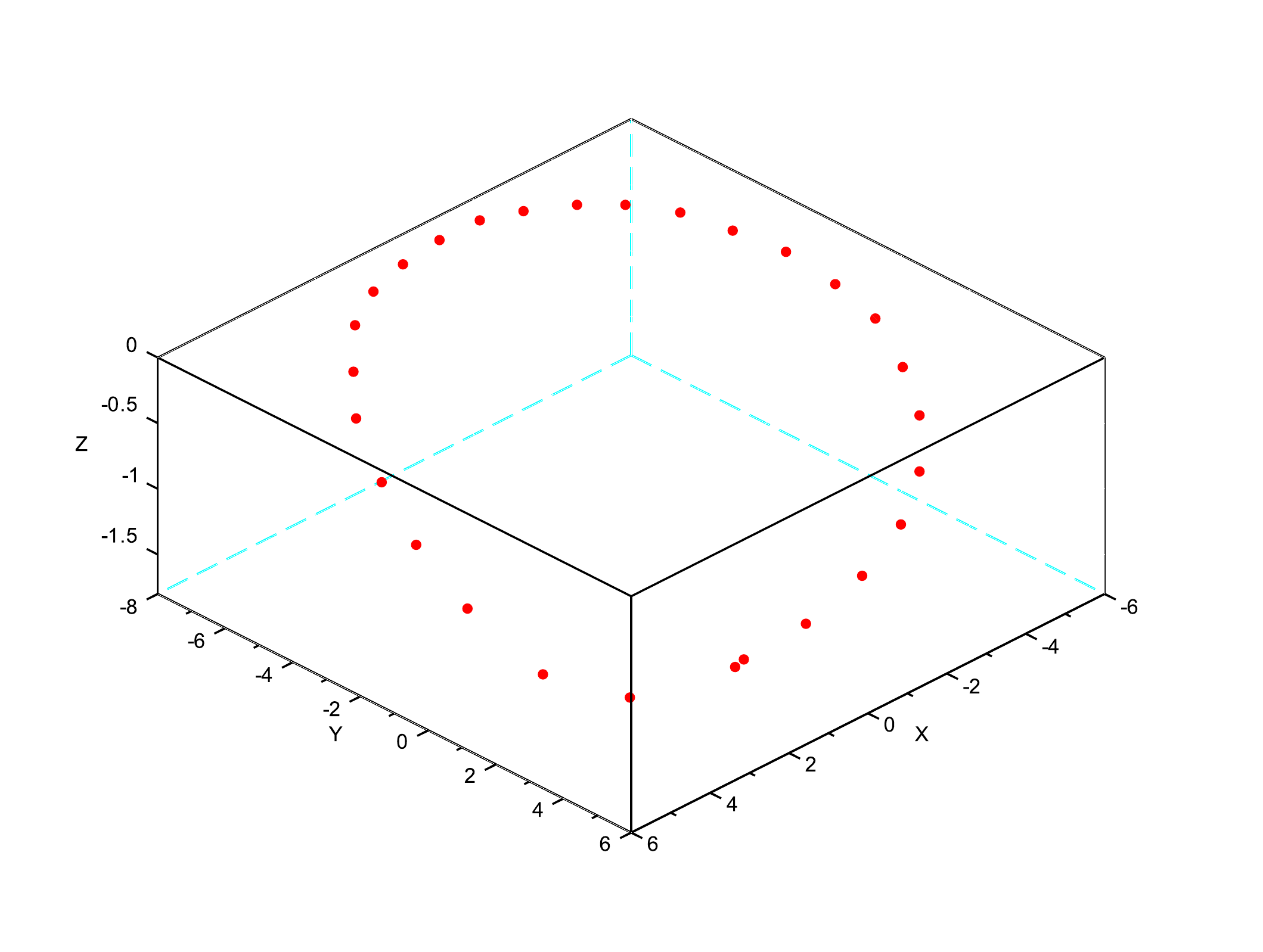}}\caption{ 
(Color Online) Scatter plot of the experimentally observed data points, in a typical dataset. One observes approximate orbit closure.}
\end{center}
\end{figure}

\begin{figure}
\begin{center}
\scalebox{0.4}{\includegraphics{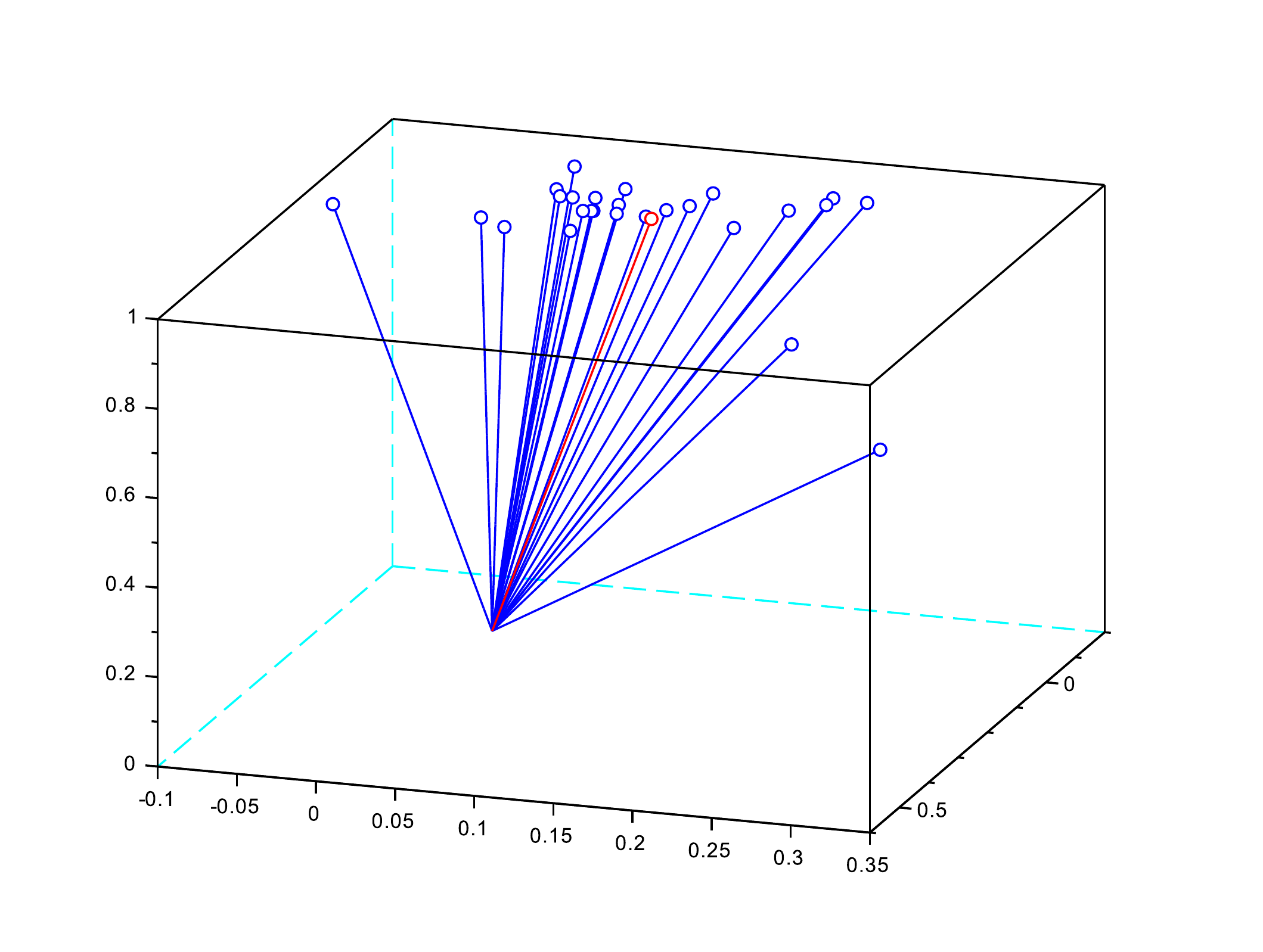}}\caption{ 
(Color Online) The distribution of the normal vectors of consecutive data points. Each pair gives a normal vector which points in a slightly different direction. For reference, the normal vector of the best fitted plane is also shown, annotated in red.}
\end{center}
\end{figure}

\begin{figure}
\begin{center}
\scalebox{0.4}{\includegraphics{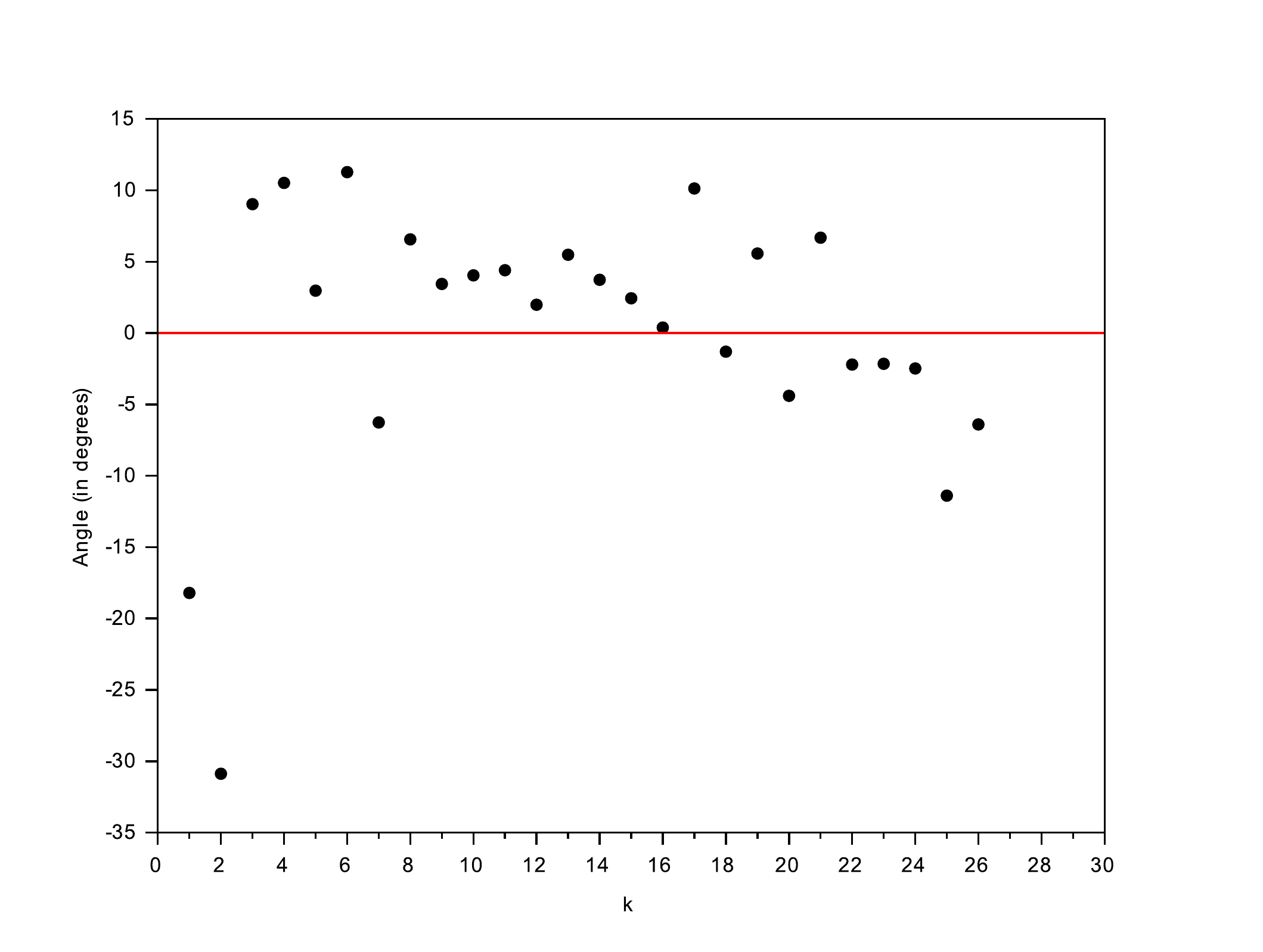}}\caption{ 
(Color Online) The distribution of the angles between consecutive normal vectors. The maximum variation between consecutive angles is $\sim 30^{\circ}$, though the typical deviation angles are small.}
\end{center}
\end{figure}

However, these results resoundingly indicate that even though the motion is strictly speaking not in a plane, the observed data points are still scattered about some best fitted plane, with marginal deviations. Thus, although this is strictly speaking, not motion in a plane, but the deviations from perfect coplanarity is small, implying therefore that these data points are not too far from being in a plane. 

In the following, we suggest a remedial procedure which may be undertaken to still approximately salvage Kepler's first law for the purpose of making a computational orbital mechanics lab, in spite of all the preceding discussion in this section.   

\end{section}


\begin{section}{Resurrecting Kepler's First Law}

\label{Resurrection_Section_KeplerLaw1}

\subsection{Best Fitting a Plane to Data}
\label{Plane_fitting_section}

\subsubsection{Principal Component Analysis}
\label{PCA_subsubsection}

Plane fitting can be achieved through the powerful technique of Principal Component Analysis (PCA) \cite{Pearson_PCA_orgnl, Jollife_PCA_book, Shlens_TutorialPCA, Bishop_PatternRecog_ML, Smith_OtagoPCAReport, SE_MakingSenseofPCA, Gnanadesikan}. In essence, this method 
relies on the identification of 
the principal components/axes for the data \cite{Gnanadesikan}, through the aid of the covariance matrix, whose each element measures the covariance/correlation between the coordinates. The PCA procedure \cite{Jollife_PCA_book, Shlens_TutorialPCA} requires this matrix to be diagonalized, and its normalized eigenvectors to be extracted and sorted according to the corresponding eigenvalues. As regards regression problems, these eigenvectors carry a special geometrical significance. For instance, for linear regression involving two variables ($x, y$), the direction (eigenvector) corresponding to the largest eigenvalue (first principal component), maximizes the variance of data projection along this component, and minimizes the residual/projection error between the original data points and their projection on to this first principal direction \cite{Bishop_PatternRecog_ML, SE_MaxVarianceAlong_MinError}
thereby effectively solving the least square fitting problem. Orthogonality of the eigenvectors \cite{Weber_Arfken} entails that the other principal direction would then give the perpendicular to this best fitting straight line \cite{Smith_OtagoPCAReport}. Plane fitting in three dimensions is 
an extension 
of the same idea \cite{Pearson_PCA_orgnl}, wherein the normalized eigenvectors corresponding to the first two principal components, identify as the basis vectors of the best fitting plane, and minimize the sum of square of projection error of all these 3D points to this best fitting plane so identified. 
The third principal component, corresponding to the smallest eigenvalue of the covariance matrix, identifies the outward normal vector ${\hat n}$ to this best fitting plane. 

Applying this technique in our context, 
we first `center' the data \cite{Pearson_PCA_orgnl}, by determining the centroid and 
co-ordinates of data points relative to 
it. 
If ${\vec r}$ and ${\vec r_{\rm C}}$ respectively denote the position vectors of any experimental data point P, and the centroid C of all $N$ data points, then the position vector of P relative to C is ${\vec R_{\rm rel}} = \left({\vec r} - {\vec r_{\rm C}} \right)$. 
In terms of its components $(X, Y, Z)$, 
we can define \cite{Shlens_TutorialPCA} the covariance between, e.g. $X$ and $Y$ as 
\begin{equation}
\sigma_{\rm XY}^{2} = \frac{1}{N} \sum_{i = 1}^N X_i Y_i, \label{correlationmatrix}
\end{equation}
Observing that $\sigma_{ij}^{2} = \sigma_{ji}^{2}$, with $\left\{ i, j \right\} \in\left\{  X,Y,Z \right\}$, and the diagonal term $\sigma_{jj}^{2}$ reduces to the usual variance $\sigma_{j}^{2}$ \cite{Weber_Arfken}, one can thus compose the symmetric, covariance matrix $\Sigma$, as
\begin{eqnarray}
\Sigma =\left[
\begin{array}
[c]{ccc}%
\sigma_{\rm X}^{2} & \sigma_{\rm XY}^{2} & \sigma_{\rm XZ}^{2}\\
\sigma_{\rm YX}^{2} & \sigma_{\rm Y}^{2} & \sigma_{\rm YZ}^{2}\\
\sigma_{\rm ZX}^{2} & \sigma_{\rm ZY}^{2} & \sigma_{\rm Z}^{2}
\end{array}
\right]
\label{Covariance_matrix}
\end{eqnarray}
While ab-initio numerical procedures for the implementation of the rest of the PCA procedure, viz. diagonalization and eigenvector-eigenvalue extraction, are well documented \cite{Jollife_PCA_book}, these can be readily performed with the aid of Scilab  \cite{Scilab_site}, with this ease being rooted in the powerful DSYEV subroutine of the ubiquitous linear algebra package LAPACK \cite{LAPACK}, thus making this procedure 
easily accessible even to novices/undergraduates.  

\subsubsection{Singular Value Decomposition}

A closely related approach, which is 
also frequently used 
in 
regression analysis
\cite{Mandel_SVD} and can also be easily adopted
here, is to construct the full Singular Value Decomposition (SVD) 
of the ${\vec R_{\rm rel}}$ obtained in Section \ref{PCA_subsubsection}. 
In general terms, the SVD of a rectangular $\left( m \times n \right)$ matrix $A$ \cite{Bisgard, Jollife_PCA_book, Strang, Atkinson} refers to the decomposition $A = U {\mathbb S} V^T$, where $U$ and $V$ are orthogonal matrices of sizes $(m \times m)$ and $(n \times n)$ respectively, and ${\mathbb S}$ is a $(m \times n)$ matrix populated only along the diagonal entries, 
bearing the `singular values' of $A$, which are simply the square roots of the eigenvalues of 
 $A A^{\rm T}$. 
Applying this SVD technique in our context, for plane fitting the centered data points, our equivalent $A$ matrix here would be a $(3 \times N)$ data matrix bearing the components of ${\vec R_{\rm rel}}$. Upon performing its SVD, the normal vector ${\hat n}$ of the best-fitting plane can be readily identified as the left singular vector (i.e. the column of $U$) corresponding to the least singular value (min$({{\mathbb S}_{\rm ii}}))$. Since $U$ is orthogonal, the other two columns of $U$ represent the null space of the vector, representing geometrically the basis vectors which span the best fitting plane. Once again, this computation of SVD of our data matrix can be greatly facilitated by Scilab \cite{Scilab_site}, wherein we can use the inbuilt DGESVD routine of the Linear Algebra package LAPACK \cite{LAPACK} to readily access the same.

The 
implementation-routine for this decomposition seems to suggest that this is an alternative take on this regression problem, but in reality, this constitutes a completely equivalent regression strategy \cite{Mandel_SVD, Shlens_TutorialPCA, SE_RelationPCASVD_1, SE_RelationPCASVD_2}. 
The PCA approach identifies (and sorts) mutually orthogonal axes as per the variance of the data along them, thereby identifying, most notably, the first principal component, viz. the special axis along which the variance of the data is the largest, or the data is clustered the most.  
Thus, the normal vector ${\hat n}$ of the best-fitting plane is, very understandably, that orthogonal axis with the lowest variance of data-projection along it. As noted before, the singular values of the matrix form of ${\vec{R}_{\rm rel}}$, are in proportion with the eigenvalues of the correlation matrix defined through Eq. (\ref{correlationmatrix}), and hence, it is only natural that identification on the basis of lowest singular value in the SVD procedure, also results in the same normal vector ${\hat n}$. Thus, in spite of appearances, both procedures are a means to the same end. Although from a computational perspective, since the creation of the covariance matrix in the PCA procedure involves square/bilinear terms, it can potentially cause precision-loss, and hence, 
instead of 
a direct identification of the lowest eigenvalue solution on the basis of PCA, the equivalent SVD approach is generally considered preferable. However, in all datasets analyzed in the present work, both these procedures are observed to yield perfectly identical results for the best fitted plane, owing to this interrelation between PCA and SVD. 

\subsection{Projection of Points on the Plane}
\label{Projection_points_plane_subsection}
While the procedures of Section (\ref{Plane_fitting_section}) identify the best fitting plane for the data points, in general, these data points do not strictly lie in this plane, but are distributed about it. Therefore, in order to formulate orbital mechanics within this plane, we must project these data points onto the plane, and consider these images (projections onto the plane) for subsequent analysis. As long as the deviation angle identified in Section (\ref{Critical_Assessment_KeplerFirstLaw_section}) is small, these images would very nearly correspond to the actual data points. In the following, we adopt the projection convention that the aforementioned `image' point is that point on the plane that is nearest to the actual experimental data point. 

In linear regression, it is a familiar fact that the best fitted straight line passes through the centroid of the data points \cite{Pearson_PCA_orgnl, Chernov_Regression_Book, Chernov_Uniqueness}. This can be readily generalized to three dimensions, and thus, one may infer that the best fitted plane, identified in Section (\ref{Plane_fitting_section}), passes through the centroid of the data points. Thus, we know both, the normal vector $\left({\hat n}\right)$ to the best fitting plane, 
and one point which surely lies on the plane. This much is sufficient information to determine the image points on the plane, in accordance with the following procedure \cite{Anton, SO_Projection, GeometricToolsBook}. 
Invoking the ${\vec R_{\rm rel}}$ from Section (\ref{PCA_subsubsection}), which represents the position vector of any experimental point P relative to the centroid, $({\vec R_{\rm rel}} \cdot {\hat n})$ gives its projection on ${\hat n}$, or the distance from point to plane along the normal. Subtracting the corresponding projection vector $({\vec R_{\rm rel}} \cdot {\hat n}) {\hat n}$ from ${\vec R_{\rm rel}}$ returns the component of ${\vec R_{\rm rel}}$ perpendicular to $\hat n$, thus lying in the plane and representing the position vector of the image point relative to the centroid. 

Then, relative to our original $(x,y,z)$ coordinate system, the position vector of the image point ($\vec r_{\rm (Im)}$) can then be constructed by subtracting the projection vector from ${\vec r}$, as
\begin{equation}
{\vec r_{\rm (Im)}} = {\vec r} - ({\vec R_{\rm rel}}\cdot {\hat n}) {\hat n} = {\vec r} - \left( \left({\vec r} - {\vec r_{\rm C}} \right) \cdot {\hat n} \right) {\hat n}
\end{equation}

\subsection{Transforming coordinates to the plane of motion: from 3D to 2D data}
\label{from3Dto2D}
Next, we seek to transform 
the original $(x, y, z)$ coordinates of these image points to new coordinates within the plane identified above. This requires a reorientation of the co-ordinate system, such that the latter coordinates may be denoted as $(x_p, y_p, z_p)$, with $z_p = 0$, and hence, we can perform the subsequent orbital mechanical analysis on the two coordinates $(x_p, y_p)$ within the plane. This transformation may be achieved through a couple of equivalent procedures. 

\subsubsection{Geometric Determination of Rotation Matrix}
\label{Geom_Det_Rot_Mat}
Mathematically, this transformation from the original to the new coordinates may be realized through rotations by three Euler angles $(\alpha, \beta, \gamma)$, eventually constituting the rotation matrix $\mathbf{R} = R_1\left({\hat z}, \alpha \right) R_2\left({\hat y}, \beta \right) R_3\left({\hat z}, \gamma \right)$ \cite{Weber_Arfken, Goldstein}. While it is usual to construct this cumulative rotation matrix, by individually working out these $R_i$ matrices, in this context, we are less concerned with the individual pieces and more with the end result $\mathbf{R}$, so an alternative procedure \cite{MSE_make2Dfrom3D} is more convenient. The transformation between the original and the projected coordinates may be represented as: 
\begin{eqnarray}
\left[
\begin{array}
[c]{c}%
x_p \\
y_p \\
z_p %
\end{array}
\right] = \mathbf{R} \left[
\begin{array}
[c]{c}%
x \\
y \\
z %
\end{array}
\right] =\left[
\begin{array}
[c]{ccc}%
e_{1}^{x} & e_{2}^{x} & e_{3}^{x}\\
e_{1}^{y} & e_{2}^{y} & e_{3}^{y}\\
e_{1}^{z} & e_{2}^{z} & e_{3}^{z}%
\end{array}
\right]
\left[
\begin{array}
[c]{c}%
x \\
y \\
z %
\end{array}
\right]
\label{Eq_projcoords}
\end{eqnarray}
Invoking the properties \cite{Weber_Arfken} of the rotation matrix ${\mathbf R}$, 
the vectors $e_{i}^{\upsilon}$ for $\upsilon \in\left\{  x,y,z\right\}  $ make an orthonormal set, satisfying
$
\sum_{i=1}^{3}e_{i}^{\upsilon_1}e_{i}^{\upsilon_2}= \delta_{\left( \upsilon_1, \upsilon_2 \right)}$. Using this, these vectors can be determined simply through vector calculus and geometric considerations. Taking any three points $P_1$, $P_2$, and $P_3$ in the plane of motion, each written as yet in terms of its $(x, y, z)$ coordinates, one can compute and normalize the vector $\overrightarrow{P_1 P_2}$, and use it as the unit vector  $e_{i}^{x}$ along the $x_p$ axis in the plane. Since this plane is to eventually be realized as the $x_p$-$y_p$  plane, unit vector $e_{i}^{z}$ along the $z_p$ axis must be perpendicular to any vectors within this plane. Since the previously computed $e_{i}^{x}$, and the vector $\overrightarrow{P_1 P_3}$, are two such vectors at hand, $e_{i}^{z}$ can be chosen to lie along the cross product of these two vectors, duly normalized. Thereafter, $e_{i}^{y}$ must be such that it is perpendicular to both $e_{i}^{x}$ and $e_{i}^{z}$, and hence, can be computed from their cross product. To summarize this geometrical procedure:
\begin{eqnarray}
\widehat{e^{x}} = \frac{\overrightarrow{P_1 P_2}}{\vert \vert \overrightarrow{P_1 P_2} \vert \vert}, \ \widehat{e^{z}} =  \frac{\widehat{e^{x}} \times \overrightarrow{P_1 P_3}}{\vert \vert \widehat{e^{x}} \times \overrightarrow{P_1 P_3} \vert \vert}, 
\ \widehat{e^{y}} =  \widehat{e^{z}} \times \widehat{e^{x}}
\label{eqn_geom_3Dto2D}
\end{eqnarray}
These equations determine ${\mathbf R}$ and the projected coordinates through the use of Eq.(\ref{Eq_projcoords}). It can be readily verified computationally that this procedure returns a $z_p \approx 0$ (typical numerical values $\sim 10^{-16}$) \cite{Github_OrbM}. One must note that since the points $P_i$ in Eq.(\ref{eqn_geom_3Dto2D}) are chosen arbitrarily out of all points in the plane, the direction of the $x_p$ axis ($\widehat{e^{x}}$) is also arbitrary within the plane. Thus, even if the $(x_p, y_p)$ data obtained through Eq. (\ref{Eq_projcoords}) represents an ellipse, it is, in general, not in the principal-axis form \cite{Anton}. 

\subsubsection{Using Basis Vectors from SVD/PCA}
\label{SVDPCA_Det_Rot_Mat}
As seen previously in Section (\ref{Plane_fitting_section}), the normal vector ${\hat n}$ of the best fitting plane, as well as two orthogonal basis vectors which span the best fitting plane, can all be directly extracted from both, the PCA and SVD approaches. These three vectors can be directly utilized to construct the rotation matrix $\mathbf{R}$, and  the projected coordinates can again be computed through Eq. (\ref{Eq_projcoords}). Unsurprisingly, this procedure similarly returns $z_p \approx 0$ (typical numerical values $\sim 10^{-16}$) \cite{Github_OrbM}.  

A few comments regarding how this compares against the procedure of Sec.(\ref{Geom_Det_Rot_Mat}) are in order. 
Since any two mutually orthogonal unit vectors in the best fitted plane can qualify as the basis vectors in that plane, an infinite number of such choices exist which correspond to the same ${\hat n}$, implying thereby that the co-ordinates  $(x_p, y_p)$ returned by the procedures of Sec.(\ref{Geom_Det_Rot_Mat}) and Sec.(\ref{SVDPCA_Det_Rot_Mat}) may disagree in general. We indeed observe this to be the case in our computations \cite{Github_OrbM}, but this totally-legitimate variation in projected coordinates is found to be inconsequential in the larger context. With either set of coordinates, the conic section identified through the procedure is found to be robust \cite{Github_OrbM} (e.g. leads to ellipses with identical eccentricities, semi-major, and semi-minor axes), in line with intuitive expectations.

\end{section}

\begin{section}{Intermission}

The above procedures turn raw experimental data into elliptical/conic section-data in  a plane, which is liable for conic section fitting, which is the logical next step. While that may be readily performed, it is natural to wonder exactly how reasonable is this resurrection of Kepler's first law? To conclusively answer this question, one measure that can be adopted is the computation of the Laplace-Runge-Lenz (LRL) vector for the motion of the marble along the elliptic orbit so defined. For Kepler orbits, this LRL vector is a constant of motion \cite{Curtis, Goldstein}, and hence, is conserved in both magnitude and direction. We analyze the same in the subsequent work \cite{OurOrbM_Art2}, and find that the same is conserved within $0.1 \%$, which may be taken as a telltale signature of the Kepler's first law being upheld near-perfectly within this formulation.

Also, since all data points ($x_p, y_p$) are separated by the same time interval $\Delta t$, dictated by the number of frames per second, of the recording camera, velocity components in the plane of the motion can be approximated through the finite difference (either of the forward difference, or the central difference) relations \cite{Weber_Arfken}. In the present work, we work with the central difference convention, and estimate
\begin{equation}
v_{x_p}(i) = \frac{\left(x_p(i+1) - x_p(i-1)\right)}{2 \Delta t}
\end{equation}
It is straightforward to verify that this is exactly how Tracker also estimates the velocity components $v_x$ and $v_y$, from the $x, y$ coordinates, within the software itself. In principle, it is also possible to import these velocity components themselves within Scilab, but then one would have to trace their transformation through all the projections and basis changes performed in Section \ref{Resurrection_Section_KeplerLaw1}, which is significantly arduous relative to the procedure outlined above. 

Thus, at this point, we possess both the ${\vec r_p}$ and ${\vec v_p}$ in the plane of motion, computed from the actual motion of the marble on the 3D printed surface. As per the well-known wisdom from classical mechanics \cite{Goldstein, Curtis}, we can then calculate all other dynamical variables from this state vector information.  

We explore the applications of this formalism, as well as some generalizations in a subsequent investigation \cite{OurOrbM_Art2}. 

\end{section}

\begin{section}*{Acknowledgments} 
This work was performed during the 2017, 2018, and 2022 editions of the annual six-week summer undergraduate research camp, \emph{Flavor of Research}, organized by the D. S. Kothari Center for Research and Innovation (DSKC), Miranda House.
The authors 
gratefully acknowledge the help from Nitnui Technology, 
Delhi, towards 3D printing the eventual $z = 1/r$ surface used seminally throughout this work, and to the in-house DSKC 3D printing facilities for 
the rest of 3D printing used in this investigation,  
which led to the final outcome.
\end{section}


\clearpage
\widetext

\begin{center}
\textbf{\large Supplemental Material: \\Computational Orbital Mechanics of Marble Motion on a 3D Printed Surface - \\ 1. Formal Basis}
\end{center}
\setcounter{equation}{0}
\setcounter{figure}{0}
\setcounter{table}{0}
\setcounter{page}{1}
\makeatletter
\renewcommand{\theequation}{S\arabic{equation}}
\renewcommand{\thefigure}{S\arabic{figure}}
\renewcommand{\bibnumfmt}[1]{[S.Ref.#1]}
\renewcommand{\citenumfont}[1]{S.Ref.#1}

\setcounter{section}{0}
\renewcommand{\thesection}{S-\Roman{section}}

\section{Calibration I - Theoretical determination of surface constants}
\begin{figure}[h]
\begin{center}
\scalebox{0.57}{\includegraphics{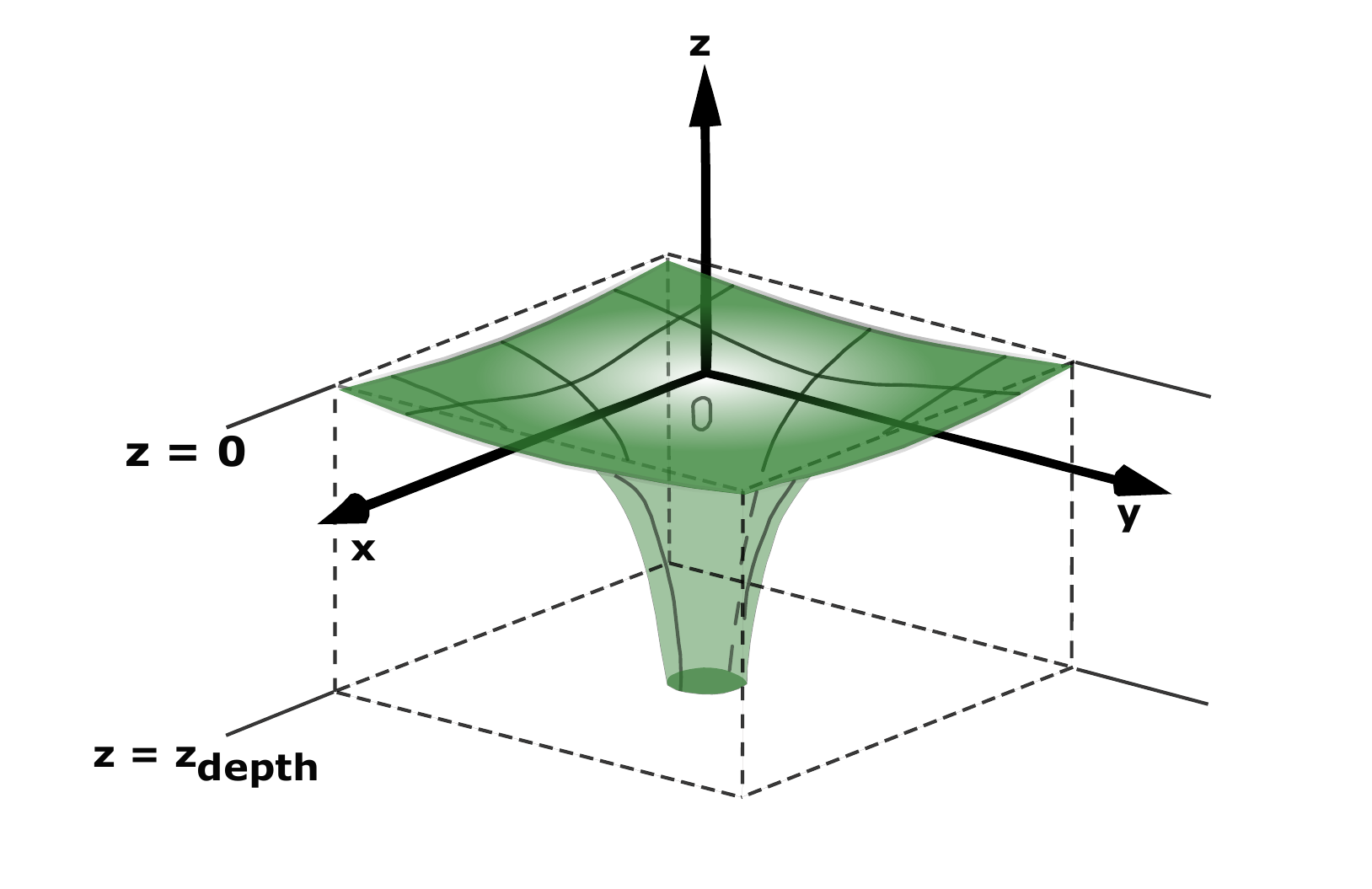}}\caption{ \label{sameas3dprinttheoryannotations} (Color Online) (\emph{Same as Fig. \ref{3dprinttheoryannotations} of the main article \cite{OurOrbM_Art1}}.) The geometry and coordinate choice for surface calibration. As before, for clarity, only the inner surface is annotated in this schematic; the complete sample region with the underlying fill is represented through dotted lines. $z=0$ is chosen at the highest elevated point of the sample, corresponding to the diagonal edges. At the bottom surface $(z = z_{\rm depth} = - \vert z_{\rm depth} \vert)$, the surface carves out a circle of radius $R_{\rm in}$.}
\end{center}
\end{figure}

\begin{figure}[b]
\begin{center}
\scalebox{0.6}{\includegraphics{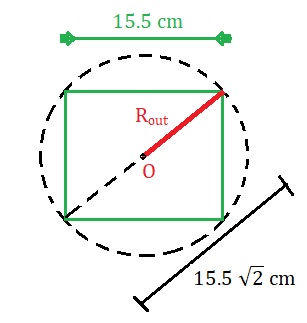}}\caption{ \label{RoutDiagram} (Color Online) An extrapolated top-lateral-view (the $x-y$ plane at $z=0$) of the sample, showing the sample circumscribed in an imagined circle of radius $R_{\rm out}$. Since the four vertices at the diagonal edges of the sample lie both in this $z=0$ plane, and on this circle, $R_{\rm out}$ can be easily evaluated in terms of the sample dimensions through Pythagoras theorem.}
\end{center}
\end{figure}

We adopt a coordinate system shown in Fig. \ref{3dprinttheoryannotations} of the main article \cite{OurOrbM_Art1}, reproduced here in this supplement as Fig. (\ref{sameas3dprinttheoryannotations}) for completeness. We choose $z = 0$ at the highest elevated points of the surface, which are the four diagonal vertices of the sample. We can sketch an $x-y$ plane through these four points, which is represented by the green square in the lateral view of the sample as shown in Fig. (\ref{RoutDiagram}). Thus, the origin of this coordinate system is located slightly above the central region of the printed surface. Additionally, this square of edge length $L_{\rm edge} (= 15.5$ cm) may be imagined as being circumscribed in a circle of radius $R_{\rm out}$ in this $z=0$ plane, such that these four vertices lie on the circle, as shown in Fig. (\ref{RoutDiagram}). It follows on geometrical grounds that this $R_{\rm out}$ can be related to $L_{\rm edge}$ through the use of Pythagoras theorem, giving $R_{\rm out} = L_{\rm edge}/\sqrt{2} = 10.96 \ {\rm cm}$. Also, from Fig. (\ref{sameas3dprinttheoryannotations}), the printed surface carves a circle of radius $R_{\rm in} = 2$ cm, at the surface $z = z_{\rm depth}$, as physically measured from the sample.


\begin{figure}
\begin{center}
\scalebox{0.7}{\includegraphics{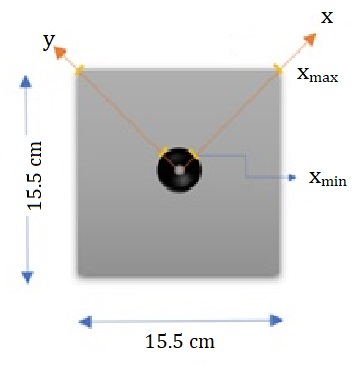}}\caption{ \label{xmaxmindiagram} (Color Online) An extrapolated top-lateral-view (the $x-y$ plane) of the physical sample, showing the choice of $x$ and $y$ axes for the process of determination of surface constants. The four vertices at the diagonal edges of the sample lie in the $z = 0$ plane, and identify $x = x_{\rm max} = R_{\rm out}$, while the inner circle edge identifies $x = x_{\rm min} = R_{\rm in}$, which occurs at $z = z_{\rm depth}$.}
\end{center}
\end{figure}
Now, with the 3D printed surface mathematically represented as 
\begin{equation}
z = C_1 + \frac{C_2}{r}, \ \ \ {\rm with} \ C_2 = - \vert C_2 \vert, 
\end{equation}
we notice that the occurrence of radial coordinate $r$ here permits rotational symmetry in the $x-y$ plane, since within this plane we may choose e.g. $x$ axis to lie anywhere.
Using this flexibility, suppose we traverse along the surface along the $y = 0$ line itself, then $r = x$ only, and in the $x-z$ plane, the surface becomes a simple curve 
\begin{equation}
z = C_1 + \frac{C_2}{x}, \ \ \ {\rm (y = 0)}
\end{equation}
Along this curve, $x$ varies from $x_{\rm max} = R_{\rm out}$ (at $z = 0$), to $x_{\rm min} = R_{\rm in}$ (at $z = z_{\rm depth} = - \vert z_{\rm depth} \vert$). This realization generates the following two equations in the two variables $C_1$ and $C_2$:
\begin{equation}
C_1 + \frac{C_2}{x_{\rm max}} = 0
\end{equation}
\begin{equation}
C_1 + \frac{C_2}{x_{\rm min}} = z_{\rm depth} = - \vert  z_{\rm depth} \vert
\end{equation}
These can be solved to obtain:
\begin{equation}
C_2 = - \vert z_{\rm depth}\vert \frac{x_{\rm max} \ x_{\rm min}}{\left( x_{\rm max} -x_{\rm min} \right)} = - \vert z_{\rm depth}\vert \frac{R_{\rm out} \ R_{\rm in}}{\left( R_{\rm out} -R_{\rm in} \right)}
\end{equation}
\begin{equation}
C_1 = - \frac{C_2}{x_{\rm max}} = \vert z_{\rm depth}\vert \frac{x_{\rm min}}{\left( x_{\rm max} -x_{\rm min} \right) } = \vert z_{\rm depth}\vert \frac{R_{\rm in}}{\left( R_{\rm out} -R_{\rm in} \right)}
\end{equation}
Substituting the values of $R_{\rm in} = 2$ cm, and $R_{\rm out} = 15.5/\sqrt{2} = 10.96$ cm, and $z_{\rm depth} = 4.756$ cm, we obtain the following theoretical estimates for the surface constants.
\begin{equation}
C_1 = 1.0616 \ {\rm cm}, \ {\rm and} \ C_2 = -11.6357 \ {\rm cm}^2
\end{equation}

\section{Transition to the experimental calibration arrangement}
\label{supp_section_experimental_calibration}

Before we describe the details of the experimental calibration, we describe the actual experimental assembly employed during calibration, and the (theoretical) surface constant values in this geometry which we later employ for comparison with the results of the experiment. 

\begin{figure}
\begin{center}
\scalebox{0.57}{\includegraphics{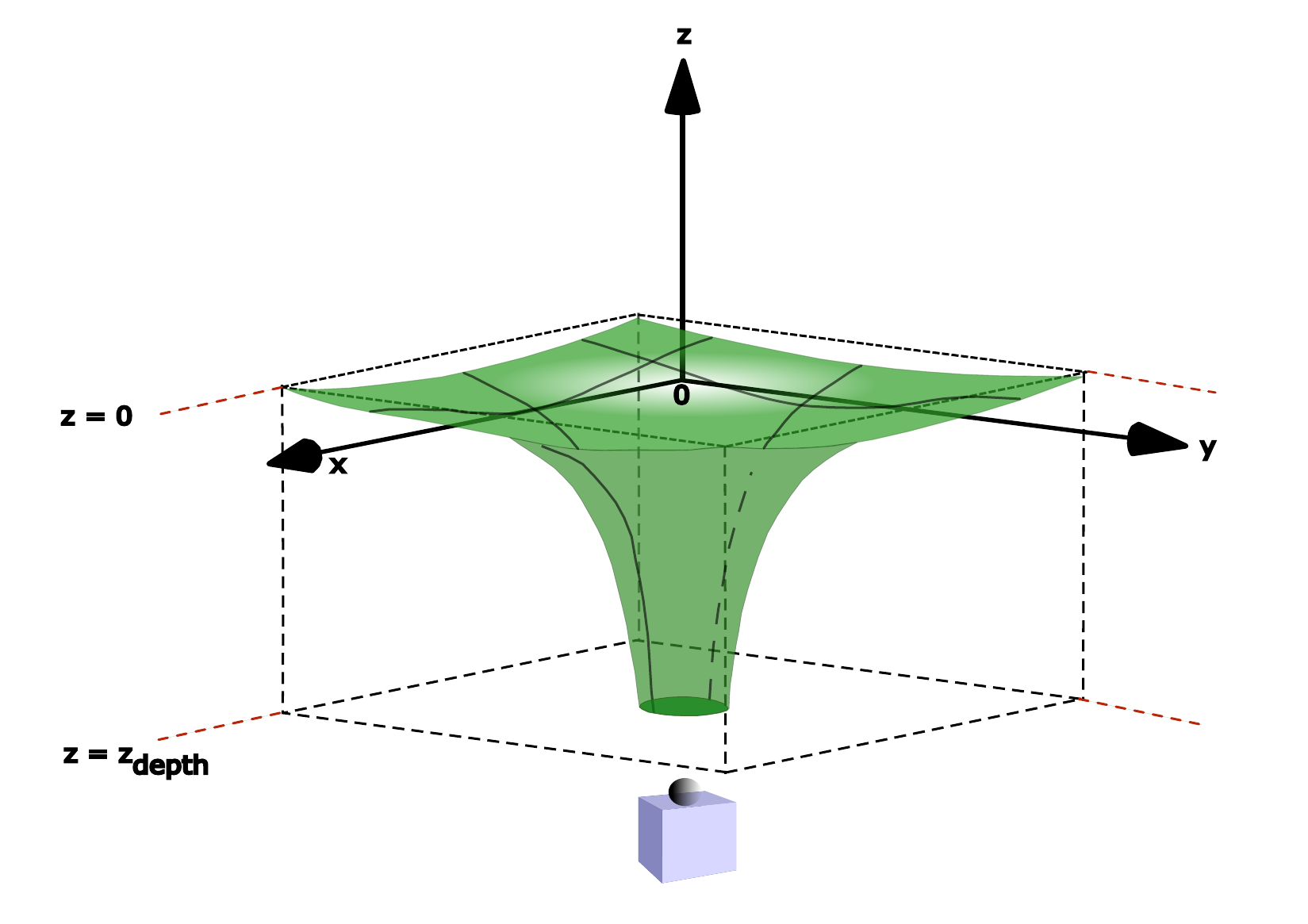}}\caption{ \label{3dprinttheoryannotations_amendedwithpin} (Color Online) A schematic diagram of the experimental setup used in calibration. As earlier, only the inner surface is annotated in this schematic; the complete sample region with the underlying fill is represented through dotted lines. The pin-thermocol arrangement is placed underneath the sample such that it lies directly below the center of the inner circle.}
\end{center}
\end{figure}

During the calibration experiment, it is useful to have a reference mark to indicate the center of the inner circle. We accomplish this elevating the sample by placing this on a hollow box with an circular orifice of appropriate size cut in the central region, and placing a marker underneath the sample, such that it coincides with the center of the said circle. The marker comprises of a small pin thrust into a piece of cuboidal thermocol plastic, as shown in Figs. \ref{3dprinttheoryannotations_amendedwithpin} and \ref{calibexpt_sslabel}. If we consider another coordinate system ($x' \ y' \ z'$) centered at the location of the pin, this new coordinate system can be related to the previous coordinate system simply through a spatial translation on the $z$ axis, while the $x,y$ coordinates remain the same between these two coordinate systems. Observing that the location of the pin would be at $(x', \ y', \ z')$ = $(0, \ 0, \ 0)$ in the latter coordinate system, and at $(x, \ y, \ z)$ = $(0, \ 0, \ - \vert z_{\rm pin} \vert)$ in the former system, the mathematical transformation between these two  coordinate frames can be written as:
\begin{eqnarray}
x' = x, \ \ y' = y, \ \ z' = z +  \vert z_{\rm pin} \vert
\end{eqnarray}
where, naturally, $\vert z_{\rm pin} \vert > \vert z_{\rm depth} \vert$, and its value can be measured. In our arrangement, $\vert z_{\rm pin} \vert = 5.356$ cm. Thus, in terms of $z'$, our $z = C_1 + (C_2/r)$ surface translates into
\begin{equation}
z' = \vert z_{\rm pin} \vert + \left( C_1 + \frac{C_2}{r} \right) = C_3 + \frac{C_2}{r},
\end{equation}
where 
\begin{eqnarray}
C_3 & = &  C_1 + \vert z_{\rm pin} \vert \nonumber\\
\ & = & \vert z_{\rm depth}\vert \frac{R_{\rm in}}{\left( R_{\rm out} -R_{\rm in} \right)} + \vert z_{\rm pin} \vert \nonumber \\
\ & \approx & 6.4179 \ {\rm cm} 
\end{eqnarray}
Thus, in this coordinate frame, the surface is represented as $z' = C_3 + (C_2/r)$, with 
\begin{equation}
C_3 = 6.4179 \ {\rm cm} , \ {\rm and} \ C_2 = -11.6357 \ {\rm cm}^2. \label{theoreticalestimatesC2C3}
\end{equation} 
We regard these as the theoretical values of the surface constants in this geometry, against which we check the experimentally obtained values of the same surface constants. 

\section{Details of Experimental Calibration}

\begin{figure}[t]
\begin{center}
\scalebox{0.55}{\includegraphics{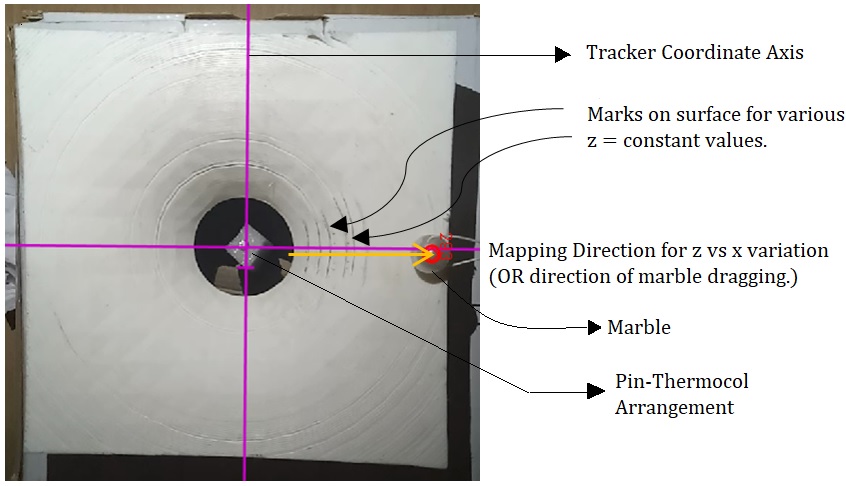}}\caption{ \label{calibexpt_sslabel} (Color Online) An annotated screen-shot of the experimental calibration with the aid of Tracker. Pencil marks were drawn on the surface for various values of $z'$ (relative to the pin underneath, regarded as $z'=0$). The $x$ axis can be digitally sketched on the surface through Tracker, and can be scaled to physical dimensions. One may read off the points of intersection through Tracker. The alternative arrangement involving marble-sliding through a string is also illustrated.}
\end{center}
\end{figure}

Adopting the assembly and geometry described in Section \ref{supp_section_experimental_calibration}, the experimental surface-calibration process requires us to sample the surface at different points in the $x-z'$ plane, and to fit a curve through these points. However, this requires a cross sectional view of the inner surface of the sample, which is not accessible for direct experimentation. With the aid of the additional reference mark provided by the pin-thermocol arrangement depicted in Fig. \ref{3dprinttheoryannotations_amendedwithpin}, we circumvent these difficulties and perform the calibration experimentally through the following procedures.

Regarding the $z'$ coordinate at the pin location as $z'=0$, initially itself, pencil marks were drawn on the surface for various values of $z'$ using a ruler held along the $z'-$ axis. (These marks may be seen on the sample in Fig. \ref{calibexpt_sslabel}.)  Thereafter, multiple different procedures can be followed for sampling the cross-sectional curve of the sample in the $x-z'$ plane. We find that the most optimal method is to once again use Tracker \cite{Tracker_Supplement}, this time to digitally sketch an $x$ axis on the surface image, with the origin chosen at the location of the pin. Using the Calibration Stick feature of Tracker, this $x-$axis can be scaled to the physical dimensions of the sample. Thereafter, we can place Tracker data marks on the points of intersection of the digital $x$ axis with the physical $z' =$ constant marks, which gives us both $x$ and $z'$ coordinates at these points of intersection. Another way of acquiring this $(x, z')$ data is to not employ a digital $x-$axis through Tracker, but to physically place a ruler on the sample, at a small distance from the center. To assist in the identification of the intersection points, we used a special marble in our possession, with a cylindrical shaft along one diameter, through which a thread could be passed and tied. By dragging this marble radially outwards on the surface through the thread, one can click pictures each time the center of the marble passes any of our constant $z'$ marks on the surface. From these images, the $x-$coordinate corresponding to the center of the marble can be read, in addition to the $z'$ value of the mark in question. Another alternative can be to use a hybrid approach -- one can make a video of the marble-dragging on the surface in a straight line, import this in Tracker, and thereafter digitally sketch an $x-$axis along this, thereby identifying intersection points of the constant $z'$ marks with the center of the marble. (This too is illustrated in Fig. \ref{calibexpt_sslabel}, with the digital $x-$axis being sketched along the horizontal. One may note that with this horizontal alignment, the $x$ coordinate does not reach the value $x_{\rm max}$ since that value is reached along the diagonal; however, in so far it proceeds, sufficient number of $(x,z')$ data points can still be generated.) After trying all three approaches, and in particular, performing extensive experimentation with the first and the third methods, we identify the first of these procedures as the best approach towards this end, while observing that the other two procedures can also generate answers with comparable accuracy but only through very careful experimentation.
 
These procedures generate a database of experimentally observed $(x_i, z_i') \ (i = 1, 2,  \ldots, n )$ values, from which the surface constants $C_2$ and $C_3$ may be obtained through simple regression procedures. Since $y = 0$ along the $x-$ axis chosen, $r = x$ itself, and we may mathematically represent the surface as $z' = C_3 + (C_2/x)$, and attempt linear regression for $z'$ against $(1/x)$, from which we can read the slope $(C_2)$ and the intercept $(C_3)$. This can be done through multiple procedures, all of them mutually equivalent:
\begin{enumerate}
\item Writing a simple routine for least squares fitting in any programming language/software package, upon importing this experimental $(x_i, z_i')$ data. 
\item Importing this data in Scilab \cite{Scilab_site_Supplement}, and using the built-in \emph{reglin} subroutine for linear regression, directly yielding the surface constants. 
\item Simply entering this experimental $(x_i, z_i')$ data in Microsoft Excel, and using Excel Solver to find an optimal fit. This can be easily done for both $z'$-versus-$(1/x)$, as well as for $z'$-versus-$x$ variation. The procedure requires some initial guesses for $C_2$ and $C_3$ to be provided, for which the theoretical estimates obtained earlier in Eq.(\ref{theoreticalestimatesC2C3}), prove very useful. The Solver optimizer searches for optimal regression solutions in the vicinity of the guesses we provide, and typically returns surface constants' values only slightly different from the theoretical values obtained earlier. 
\end{enumerate}

\begin{figure}[b]
\begin{center}
\scalebox{0.85}{\includegraphics{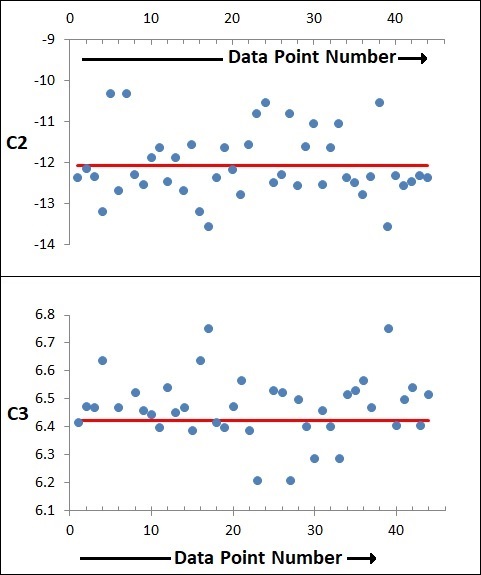}}
\caption{ \label{scatterplots_experimental_C2C3} (Color Online) Scatter plots for the experimentally determined numerical values of surface constants $C_2$ and $C_3$, across our best 44 experimental calibration datasets. The red lines indicate the respective average values over these datasets, viz. $C_2 = -12.0717 \ {\rm cm}^2$ and $C_3 = 6.42085 \ {\rm cm}$.}
\end{center}
\end{figure}

However, experimental determination of any variable may also be subject to various measurement errors. Therefore, for seeking robust determination of the surface constants, one needs to repeat the experiment several times, and assess the variation in the values obtained for the surface constants. The scatter plots of Fig. (\ref{scatterplots_experimental_C2C3}) show the individual values of surface constants determined in a total of 44 attempts at this experiment. Unsurprisingly, one finds that the surface constants obtained each time are different, with a reasonable amount of variation, but one does not identify any systematic error in the experimental findings. Thus, we simply consider the average of all the experimentally obtained values, and identify these as the experimental values of the surface constants: 
\begin{equation}
C_3 = 6.42085 \ {\rm cm} , \ {\rm and} \ C_2 = -12.0717 \ {\rm cm}^2,  \label{experimentalestimatesC2C3}
\end{equation}
\vspace{-0.7cm}
\begin{equation}
C_1 = C_3 - \vert z_{\rm pin} \vert = 1.0646 \ {\rm cm}
\label{experimentalestimateC1}
\end{equation}
Thus, we find that the theoretical and experimental values of these surface constants are in reasonably good agreement. The experimental value of $C_3$ in particular is in excellent agreement with the theoretical estimate, while the experimental value of $C_2$ is also found to be within $4 \%$ of the theoretical estimate. This agreement lends a lot of credence to this model formulated. 

As small as the difference between the theoretical and experimental estimates may be, one requires unique values of the surface constants for the computational analysis. For this, we adopt the experimental values of these surface constants, given by Eqs. (\ref{experimentalestimatesC2C3})-(\ref{experimentalestimateC1}), as the surface constant values to be used consistently for the entire computational analysis of the marble motion on this surface.

\end{document}